\documentstyle[aps,prl,twocolumn,epsfig,floats]{revtex}
\newcommand{\ua}{\uparrow}
\newcommand{\da}{\downarrow}
\newcommand{\ra}{\rangle}
\newcommand{\bfs}{\mbox{\bf S}}
\begin{document}
\draft
\twocolumn[\hsize\textwidth\columnwidth\hsize\csname @twocolumnfalse\endcsname
\title{Exact ground state and kink-like excitations of a two dimensional
 Heisenberg antiferromagnet}
\author{Rahul Siddharthan}
\address{Department of Physics, Indian Institute of Science, Bangalore 560012,
         India}
\date{\today}
\maketitle

\begin{abstract}
A rare example of a two dimensional
Heisenberg model with an exact dimerized ground state is
presented. This model, which can
be regarded as a variation on the kagom{\'e} lattice, has several features
of interest: it has a highly (but not macroscopically) degenerate ground
state; it is closely related to spin chains studied by earlier
authors; in particular, it exhibits domain-wall-like
``kink'' excitations normally associated only with one-dimensional systems.
In some limits it  decouples into non-interacting chains, purely
dynamically and not because of weakening of interchain couplings: indeed,
paradoxically, this happens in the limit of strong coupling of the
chains.
\end{abstract}
\bibliographystyle{unsrt}

\vskip2pc] 

\begin{figure}
\epsfig{file=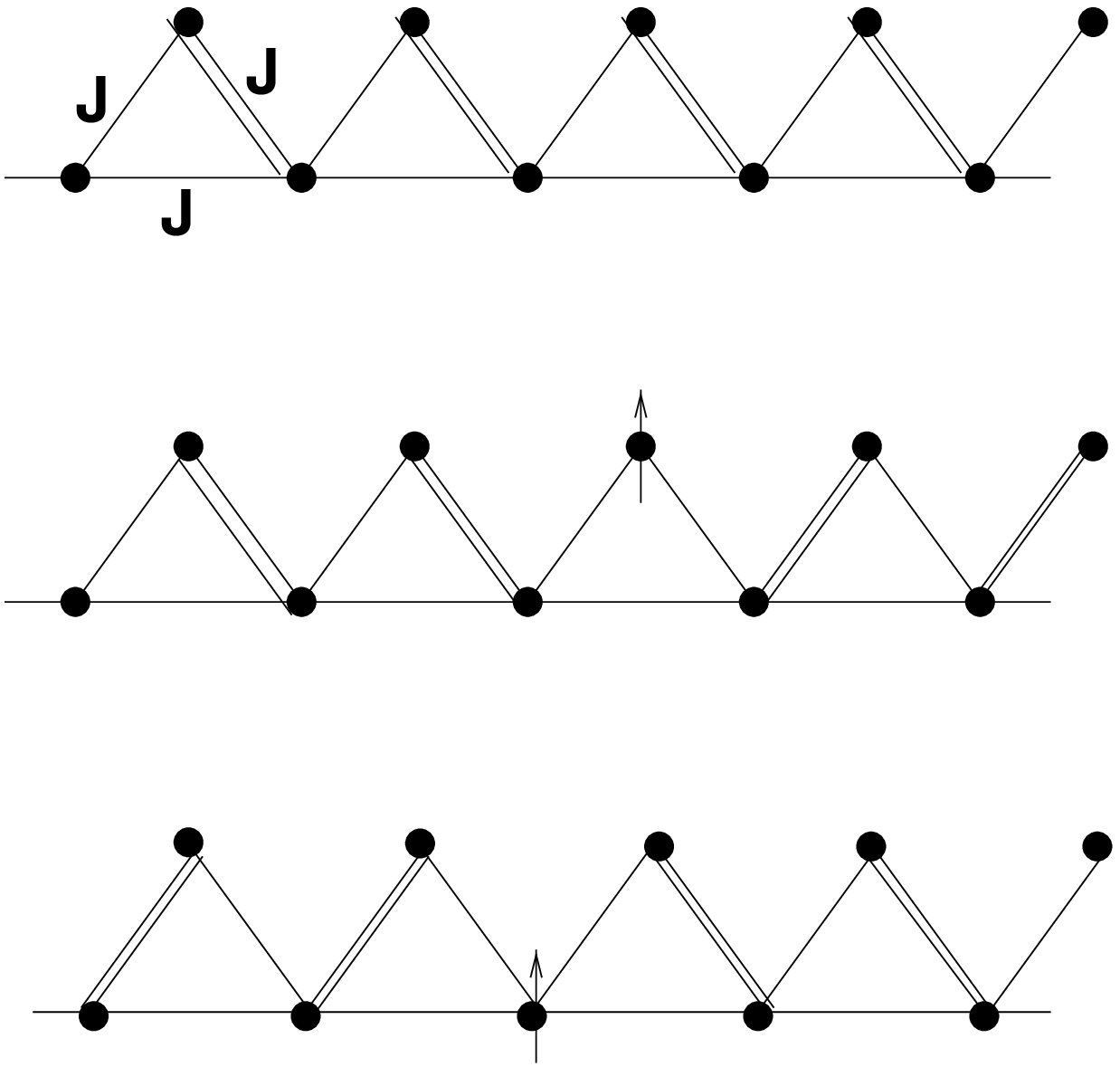, width=3.3cm} 
\hfill
\epsfig{file=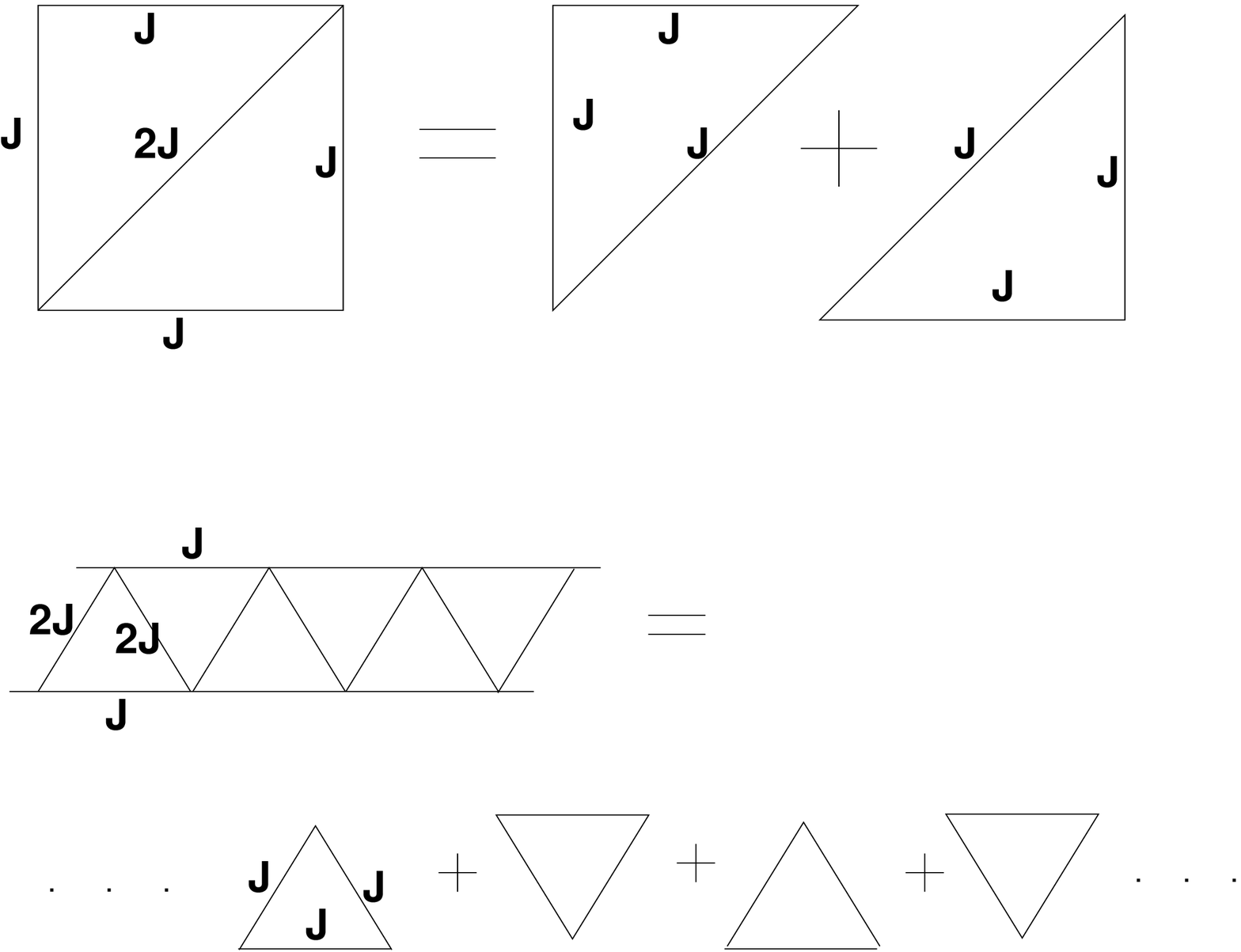, width=4.3cm}
\vspace*{1.5ex}
\caption{\label{saw}
(left) The ``sawtooth lattice'', a set of corner-sharing triangles with 
exchange $J$ along each side, two dimerized ground states, and
two distinct kinds of domain-wall-like excitations.
}
\vspace*{1.5ex}
\caption{\label{square}
(right) Above, a square as a sum of triangles; below, the Majumdar-Ghosh
chain as a sum of triangles.
}
\end{figure}

Exact many-body solutions are rare in physics. ``Integrable systems'',
systems which have as many integrals of motion as degrees of freedom
and can in principle be solved exactly, are much sought after and
nearly all the interesting examples are one dimensional.  Even the
more modest goal of obtaining an exact ground state for a non-trivial
many body problem is not easy.  The value of an exact ground state is
obvious in studying the low temperature physics. Nevertheless, in the spin
half Heisenberg model of magnetism, for instance, very few exact
ground states (other than Bethe's famous solution of the
nearest-neighbour chain 
\cite{bethe}) are known in the antiferromagnetic case:
notable examples are mostly quasi-one-dimensional, such as the
Majumdar-Ghosh chain \cite{majumdar} and the
sawtooth lattice \cite{sen,kubo} (fig.\ \ref{saw}). These have doubly
degenerate, dimerized ground states, and consequently, localized
domain-wall-like excitations. In two dimensions, this author knows of only
two exact
solutions. One is by Shastry and Sutherland
\cite{shastry} of a square lattice with alternating diagonal bonds;
an experimental
realization has recently been found \cite{uedaa} and the model
extended to three dimensions \cite{uedab}. The other appears in
a paper whose main thrust is something else (chiral terms and
three-spin interactions \cite{senchitra}). Both models have
non-degenerate ground states and excitations consist of breaking
of singlet pairs; unlike in the 1D systems, there are no 
propagating domain walls.

\begin{figure}
\epsfig{file=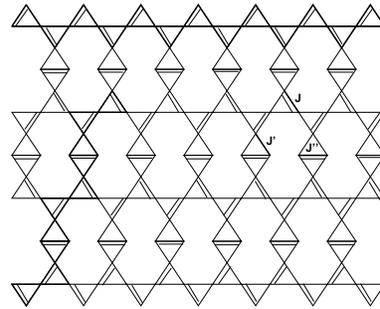, width=5cm}
\caption{\label{chains}
The lattice described in this paper, with an exact dimerized ground state
($2^L$-fold degenerate, where $L$ is the number of sawtooth-like chains)
and domain-wall-like excitations. The ground state is as described if
$J'' > 1.41 J'$. If, in addition, $J' \gg J$, the excitations are mainly
confined to the horizontal sawtooth-like chains like the one highlighted
on top by bold lines. For $J$, $J'$, $J''$ comparable, we also get
excitations in the vertical direction, which can propagate for instance
along the chain shown with bold lines at the bottom left.
}
\end{figure}

Here I present what is, as far as I know, only the third 
example of a two dimensional spin half
Heisenberg antiferromagnet with an exact ground state, one which is
quite different from the two above. The lattice is
shown in figure~\ref{chains}; it bears some visual similarity to the
much studied kagom{\'e} lattice, to which it reduces if we collapse the
diamond-shaped plaquettes (for instance by introducing a strong
ferromagnetic interaction between the extreme corners). But its
interest arises from the facts that (a)~it has exact, dimerized ground
states like the Majumdar-Ghosh and sawtooth chains, (b)~ the ground
state is highly degenerate (though not macroscopically so), and
(c)~the low energy excitations are domain-wall-like,
connecting different ground states.  This is thus the first genuinely
two dimensional spin system to display these properties. These happen
because it is highly anisotropic, and decouples in some limits into
essentially non-interacting sawtooth-like chains; this arises from
energy considerations and not from weakening of inter-chain couplings.

All the models mentioned above (except the nearest-neighbour chain)
have dimerized ground states, consisting of pairs of spins in the
singlet (zero-spin) state. Moreover, they all have the property that
the Hamiltonian is a sum of smaller Hamiltonians each of which has an
exact dimerized solution, and the full solution is constructed of
these. The general idea of constructing exact solutions piecewise 
is not new \cite{suthshastry,affleck}
but given the direct importance it has in our problem it is worth 
showing explicitly for these examples.
Consider the Heisenberg antiferromagnet Hamiltonian
\begin{equation}
H = \sum_{i,j} J_{ij} \mbox{\bf S}_i \cdot \mbox{\bf S}_j
\end{equation}
where $J_{ij}$ are positive constants. The Hamiltonian can be written
in terms of the Pauli matrices as
\begin{equation}
H = \sum_{i,j} \frac{J_{ij}}{4} \left[ 2\left(\sigma^+_i \sigma^-_j + 
              \sigma^+_j \sigma^-_i \right) + \sigma^z_i \sigma^z_j \right]
\end{equation}
where $\sigma^\pm$ $=$ $(\sigma^x \pm \sigma^y)/2$ and $\mbox{\bf S}$
$=$ $\frac{1}{2}\left(\sigma^x,\sigma^y,\sigma^z\right)$.  
It is easily verified that (a)~the ground state of the two atom
chain is a singlet (spin zero), (b)~the (four-fold degenerate) ground state
of the three atom ring is a singlet pair and a ``free'' spin.

The sawtooth chain is a chain of such triangles, joined at the
corner. Thus in the ground state, each triangle has one
side whose spins form a singlet, and the ``free spin'' of each 
triangle is part of a singlet pair on the next triangle 
(fig.~\ref{saw}). In all
the figures, a double line joining two sites indicates a 
singlet pairing of those spins.

Consider, next, a four-spin system made by combining two of these
triangles: the Hamiltonian of this system is the sum of two triangle
Hamiltonians, which is equal to a square with sides of strength $J$
and a diagonal exchange of strength $2J$. A possible ground state of
this system is a dimer along this diagonal. Another ground state is a
pair of dimers along opposite sides. For a diagonal exchange $J'$
different from $2J$ the former remains an eigenstate, and in fact it
is the ground state for $J'$ is more than roughly $1.41 J$. 
The total spin of the diagonally-connected pair is conserved: labelling
these as $\bfs_1$ and $\bfs_2$, and the other two as $\bfs_3$ and 
$\bfs_4$, the
Hamiltonian is $J'\bfs_1\cdot\bfs_2$ $+$ $J(\bfs_1+\bfs_2)\cdot(\bfs_3+\bfs_4)$
which commutes with $(\bfs_1+\bfs_2)$.   

The Majumdar-Ghosh chain can be regarded as a chain of such
side-sharing triangles (fig.\ \ref{square}). Again, the Hamiltonian is a
sum of triangle-like Hamiltonians, and the dimerized states are ground
states for each of these individual Hamiltonians---hence for the whole
system.

\begin{figure}
\centerline{\epsfig{file=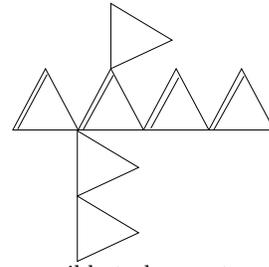, width=3.5cm}}
\caption{\label{imposs}
It is impossible to have a two dimensional lattice of corner-sharing
triangles (such as the kagom{\'e} lattice) with an exact dimerized
ground state. For such a lattice must have intersecting dimerized
sawtooth chains, but two intersecting sawtooth chains with a common
triangle and periodic boundary conditions cannot both be fully
dimerized.
}
\end{figure}

To extend this sort of model to two dimensions is another matter. 
The only published examples this author knows of are the two mentioned
earlier, the square lattice
with alternating diagonal interactions \cite{shastry,uedaa,uedab}
and the model of Sen and Chitra \cite{senchitra}, both of which can
again be built up of these elementary pieces.
By analogy with the sawtooth chain of corner-sharing triangles, the
kagome lattice may seem to be a candidate, but such ``dimerization''
is not possible on it, nor on any two dimensional lattice of corner
sharing triangles.  To see this, note (fig.\ \ref{imposs}) that any
two-dimensional dimerized network of corner-sharing triangles must
contain two infinite chains of triangles (or closed loops for a
periodic lattice), containing one triangle in common; but if one chain
is fully dimerized, it is impossible to fully dimerize the other.
In particular one cannot ``deplete'' a kagom{\'e} lattice to
obtain an exact dimerized ground state, while retaining its 2D
character. One can, of course, deplete the kagome lattice in such a
way as to destroy its two-dimensional character, and then a dimerized
ground state is possible \cite{hooley}.

With this background, consider the lattice in figure~\ref{chains}.
This lattice is superficially similar in appearance to the kagome
lattice, which can be viewed as a set of connected sawtooth chains:
but by introducing the additional rhombuses between the chains, with
interactions $J'$ along the sides and $J''$ along the short diagonals,
where $J'' >
1.41J'$ roughly, we obtain a system with an exact ground state. This is a
state where the sawtooth-like chains are dimerized as usual, while the
connecting rhombuses are dimerized along their short diagonals. $J'$
must be sufficiently small compared to $J''$ for the dimerization of
the rhombus to be its ground state, but is otherwise arbitrary, and $J$
is arbitrary compared to both of these.  The ground state of such a
system with periodic boundary conditions has a degeneracy $2^L$, where
$L$ is the number of sawtooth-like chains, each such chain being
doubly degenerate.

In the limit $J \gg J',  J''$, of course, the
system decouples into noninteracting sawtooth chains; but the interesting
thing is that the same thing happens, effectively, even when
$J'' \gg J' \gg J$, which would appear to be a strong coupling
limit between the chains. For in this case it is expensive to break
the diagonal singlet pairs in the rhombuses, so they tend to remain in
their ground states, and the spin dynamics becomes confined within
each chain. In both these limits, there is little to add in the
treatment of this lattice to the discussion of the sawtooth chain.
To recapitulate the discussion in \cite{sen}, there are two kinds of 
domain-wall excitations, which can be dubbed ``kinks'' and
``antikinks'', of which the ``kinks'' are exact eigenstates with zero
energy, while the ``antikinks'' are not exact eigenstates and have
a non-trivial dispersion with a gap. If we write a momentum eigenstate
using antikinks as follows:
\begin{equation}
|k\rangle = \frac{1}{N} \sum_{n=1}^N e^{ikn} |n\rangle
\end{equation}
where $|n\rangle$ is the state with an antikink on the $n$th
triangle, 
we can estimate the energy of such a state by calculating the
expectation $\langle k| H| k\rangle/\langle k|k \rangle$. 
This yields a dispersion 
\begin{equation}
E = (5/4 + \cos k) J
\end{equation}
derived by Sen {\em et al.\ }\cite{sen}. 
To get a better estimate they consider more
states, and show that the correction isn't very large. In particular,
the system has a gap of around $0.25 J$. 
In a periodic system the kinks and antikinks must exist in pairs. So
the system as a whole has a gap, and at low temperatures, travelling
kink-antikink excitations. In our two dimensional system, if
$J'' > 1.41J' \gg J$, the system will consist of effectively
non-interacting sawtooth chains with horizontal kink-antikink
excitations but, at low temperatures, no excitations travelling
in the ``vertical'' direction.

Things are less simple when $J$, $J'$, $J''$ are comparable. Then
excitations can propagate in the ``vertical'' direction too.
Particularly interesting is the choice $J' = J$, $J'' = 2J$: 
in this case the Hamiltonian can be written as a sum of triangles,
with the exchange interaction $J$ along each ``side''.
It is clear that excitations with energy of at
least order $J$ should exist, and this is not too far away from the
sawtooth-chain gap of $0.25J$. We improve on this below.

While the ``horizontal'' chains are disjoint in that they have no
common sites, and can be treated individually, this is not true of the
``vertical'' chains, which crisscross and intersect heavily.  In other
words, it would not be reliable to treat the vertical excitations as
excitations of vertical chains. We instead present a very rough
calculation (which can be treated as a liberal upper bound only) of
the dispersion of an excitation in one such vertical chain. 

\begin{figure}
\epsfig{file=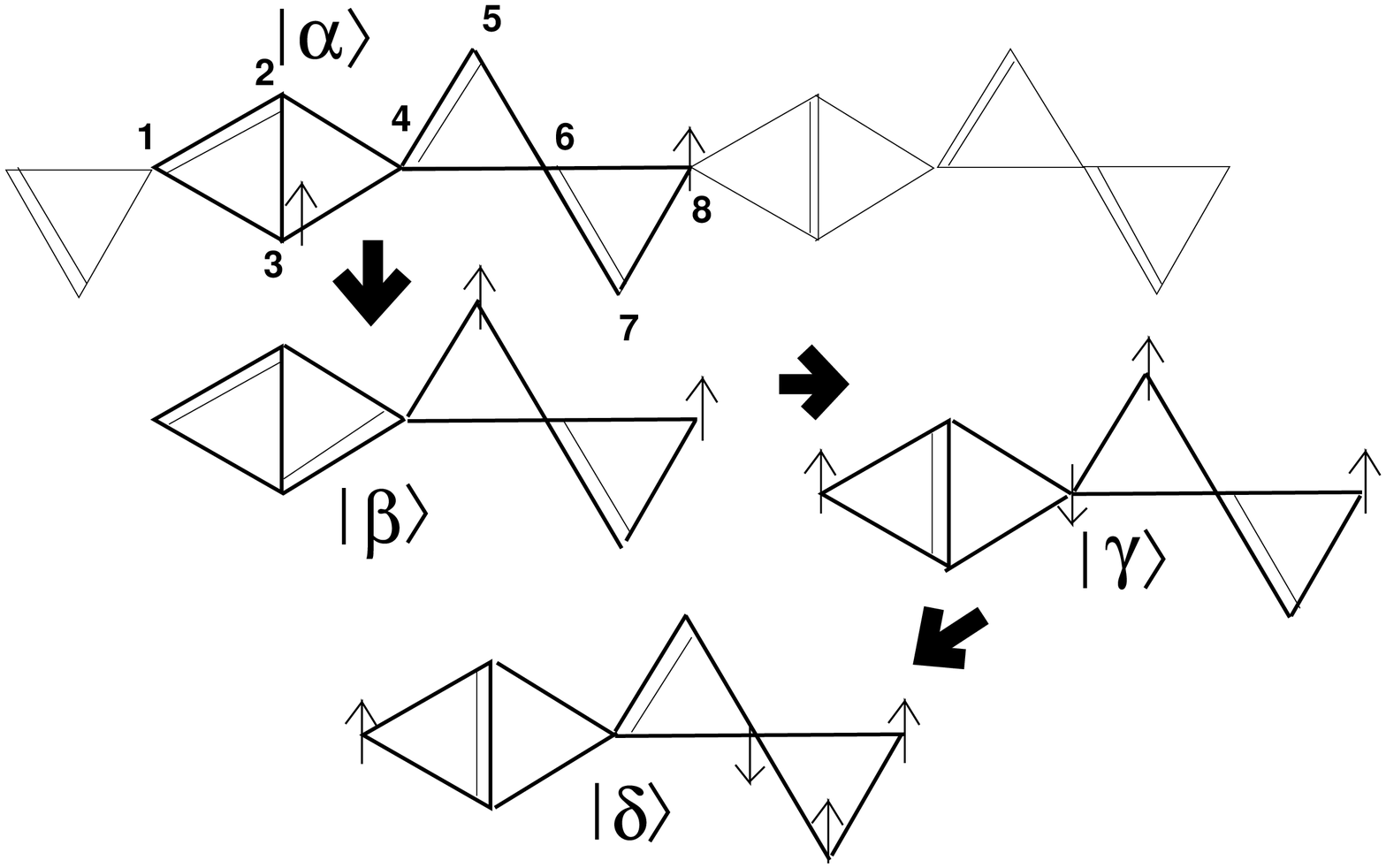, width=7cm}
\caption{\label{chainstates}
Propagation of excitation along vertical chain: Four basis
states used in variational estimate of energy}
\end{figure}

We write the Hamiltonian as a sum of triangle terms,
\begin{equation}
H = \sum_l H_l,
\end{equation}
where each individual triangle Hamiltonian is a sum of spin-spin
interactions over each side of a triangle
\begin{equation}
H_l = J \left( \mbox{S}_{1l} \cdot \mbox{S}_{2l}  +
               \mbox{S}_{2l} \cdot \mbox{S}_{3l} + 
               \mbox{S}_{3l} \cdot \mbox{S}_{1l} \right) + \frac{3}{4} J.
\end{equation}
The constant piece $3J/4$ changes the energy of a dimerized triangle
(therefore also the ground state energy) to zero.
The subscript $l$ denotes the $l$-th triangle. 
The propagation of the excitation along a chain may be regarded as 
occurring as shown in figure~\ref{chainstates}. A single non-dimerized
triangle travels through the chain via rearrangements of singlet
bonds. (Incidentally, the chain itself is another example of a
spin chain with an exact dimerized ground state. Since each pair of
corner-sharing triangles can exist in one of two states independently
of the rest of the chain, from which it is separated by a rhombus, the
ground state is macroscopically degenerate.) 
The excitations we consider 
are not domain-wall excitations. Since the corner-sharing triangles
here are part of the sawtooth chains in the full lattice, it would be
expensive for a propagating domain wall to disturb them, so we assume
that they remain unchanged except at the sites of the excitation.
We use a variational calculation with a momentum
eigenstate formed from the four basis wavefunctions shown in
figure~\ref{chainstates}:
\begin{equation}
|k\rangle = \sum_{n=1}^N e^{ikn}\left( |n\alpha\rangle +
  \beta|n\beta\rangle  + \gamma|n\gamma\rangle  
      + \delta|n\delta\rangle  \right),
\end{equation}
where the sum is over units of the sort shown in
fig.~\ref{chainstates}, and there are $N$ such units, for each of which
we assume one of the four basis states in fig.~\ref{chainstates}. 
$|n\alpha\rangle$ means the $n$th unit has configuration $|\alpha\rangle$,
and so on. The wavefunction is orthogonal to the ground state in the
$N \rightarrow \infty$ limit.
There are, of course, many more possible basis states, but
the calculation grows tedious and our purpose, which is to demonstrate
that low-energy excitations along such chains can exist, will be
served with this wavefunction.

The above basis states can be written 
\begin{eqnarray}
|\alpha\ra & = & |(\ua\da-\da\ua)\ua(\ua\da-\da\ua)(\ua\da-\da\ua)\ua\ra \\
|\beta\ra & = & |(\ua\da-\da\ua)(\ua\da-\da\ua)\ua(\ua\da-\da\ua)\ua\ra \\
|\gamma\ra & = & |\ua(\ua\da-\da\ua)\da\ua(\ua\da-\da\ua)\ua\ra \\
|\delta\ra & = & |\ua(\ua\da-\da\ua)(\ua\da-\da\ua)\da\ua\ua\ra 
\end{eqnarray}
where the ordering of the spins is as shown in
figure~\ref{chainstates}, and one should also include a normalizing
factor of $1/\sqrt 2$ for each ``singlet pair''. Our dispersion
with this state will be
\begin{equation}
 E(k) = \langle k |H|k \rangle / \langle k | k \rangle
\end{equation}
where
\begin{eqnarray}
\langle k | k \rangle & = & \frac{1}{N} \sum_{m=1}^N \sum_{n=1}^N
   e^{ik(n-m)} \left[ \langle m\alpha| + \beta^*\langle m\beta| + \gamma^*
        \langle m\gamma| \right. \nonumber \\
 & + & \left. \delta^*\langle m\delta| \right] 
 \times  \left[ |n\alpha\ra + \beta|n\beta\ra + \gamma|n\gamma\ra 
                + \delta|n\delta\ra \right]. 
\end{eqnarray}
We need to know several matrix elements between the basis states to
work this out, but the calculation is not hard. The result, for
large $N$, is
\begin{eqnarray}
\langle k|k \rangle &=& \frac{3}{4} + \left( \frac{15}{16} -
   \frac{1}{4\sqrt{2}}\right)|\beta|^2 + \left(\frac{1}{2} -
   \frac{1}{4\sqrt{2}}\right)|\gamma|^2 \nonumber \\
   & & + \frac{1}{2}|\delta|^2
   -\frac{3}{4}\mbox{Re}\beta + \frac{1}{4\sqrt{2}}|\beta-\gamma|^2
\end{eqnarray}

To evaluate $\langle k|H|k\rangle$, we write $H = \sum H_l$ and
note that $\langle n,\phi|H_l|m,\psi\rangle$ (where $\phi, \psi$ =
$\alpha$, $\beta$, $\gamma$, $\delta$) is zero unless
$l=m=n$ (since $H_l$ acting on a dimerized triangle is zero) and
even for $l=m=n$ the matrix elements will exist only for
$\phi$=$\psi$; $\phi=\beta$, $\psi=\gamma$; or $\phi=\gamma$, 
$\psi=\beta$.  This gives 
\begin{eqnarray}
 \langle k|H|k \rangle &=& \frac{J}{4} \left[ 3 + 3|\beta|^2 + 2|\gamma|^2
        + 2 |\delta|^2 \right. \nonumber \\
   & & \left. -\frac{1}{\sqrt{2}}\left(\beta^*\gamma+\gamma^*\beta
     \right) \right] \nonumber \\
  &=& \frac{J}{4}\left[ 3 + \left(3 - \frac{1}{\sqrt{2}}\right) |\beta|^2
        + \left(2 - \frac{1}{\sqrt{2}}\right) |\gamma|^2 \right. \nonumber \\
  & & \left. +  2| \delta|^2 + \frac{1}{\sqrt{2}}|\beta-\gamma|^2 \right]
\end{eqnarray}

Now we need to minimize $\langle k|H|k\rangle / \langle k|k\rangle$.
We have five real parameters to vary, since there are three complex
parameters but $\delta$ appears only as an absolute square, and minimizing
on a computer gives
\begin{eqnarray}
\beta & = & -1.3522, \\
\gamma & = & -0.4781, \\
\delta & = & 0, \\
E(k)  & = &  0.59 J. 
\end{eqnarray}
This is a dispersionless excitation, but that may change with a more
careful treatment. It is also interesting that $|\delta\rangle$ does not
appear in the minimum energy wavefunction, but that too may change if we
include more basis states. The important point is that the energy 
is not too far from  the gap of the sawtooth chain ($0.25 J$) and this
estimate will certainly reduce further if we
include more basis states and account for the crossings among
these ``vertical chains'' (which intersect, unlike the horizontal sawtooth
chains). So for $J''=2J$, $J'=J$, we can expect such excitations to be
present at low temperatures together with the sawtooth-chain excitations.

A final interesting point is that the system has an infinite (but not
complete) set
of conserved quantities, namely the total spins $\bfs_D$ $=$ $\bfs_b$
$+$ $\bfs_b$ 
 along the short diagonals of
the rhombuses where $\bfs_a$ and $\bfs_b$ are the spins on the sites
connected by these short diagonals. 
All eigenstates of the system, and of the vertical
chain discussed above, can be chosen to be eigenstates of these 
$\bfs_D$: but these will not be momentum eigenstates (since
the $\bfs_D$ do not commute with the translation operator).

To summarize, the spin half Heisenberg model on the two dimensional
lattice described here has several interesting features such as
an exact dimerized ground state; a large ground
state degeneracy (exponential in the square root of the system size);
a decoupling into effectively noninteracting spin chains, which is
dynamic and not because of weakening of inter-chain coupling; and
domain-wall excitations of the kind normally found only in
one-dimensional spin chains.  The system is, paradoxically, most
one-dimensional at high inter-chain couplings $J'' > 1.41 J' \gg J$,
and at these values it is clear that the system is gapped, because the
horizontal sawtooth-chain excitations are known to be gapped. At the
other extreme, $J \gg J', J''$, the excitations are confined to the
diamond-shaped plaquettes and cannot propagate, so again the system is
gapped (the gap being of order $J''$). At intermediate values there are 
both horizontal (intra-chain)
and vertical (inter-chain) excitations and it is not certain whether the
vertical ones are gapless. Several related systems---sawtooth
chains, kagome lattices, even the Shastry-Sutherland square
lattice---have experimental realizations, and it would be interesting
to look for experimental examples of this system too.

I am grateful to Diptiman Sen and Sriram Shastry for useful
discussions.



\end{document}